\documentclass{aa}
\usepackage{psfig}
\newcommand\aap{{A\&A}}

\newcommand\apjl{{ApJ}}

\newcommand\mnras{{MNRAS}}
\newcommand\nature{{Nature}}

\def\spose#1{\hbox to 0pt{#1\hss}}
\newcommand\simlt{\mathrel{\spose{\lower 3pt\hbox{$\mathchar"218$}}
     \raise 2.0pt\hbox{$\mathchar"13C$}}}
\newcommand\simgt{\mathrel{\spose{\lower 3pt\hbox{$\mathchar"218$}}
     \raise 2.0pt\hbox{$\mathchar"13E$}}}

\begin{document}

\thesaurus{09 (09.13.2; 09.03.1; 09.04.1)}

\title{Dehydrogenation of polycyclic aromatic hydrocarbons in the diffuse interstellar medium}

\subtitle{}

\author{My Ha Vuong\thanks{Present address: Observatoire de la C\^ote d'Azur, D\'epartement Fresnel, F-06304 Nice Cedex, France} \and Bernard H. Foing}

\offprints{Bernard H. Foing}

\institute{Solar System Division, ESA Space Science Department, ESTEC/SO, 2200 AG Noordwijk, The Netherlands; vuong@obs-nice.fr, bfoing@estec.esa.nl}
	
\date{Received 2000 ; accepted 2000}
\authorrunning{Vuong \& Foing}
\titlerunning{Dehydrogenation of PAHs}
\maketitle

\begin{abstract}   

We present a model for the hydrogenation states of Polycyclic Aromatic Hydrocarbons (PAHs) in the diffuse interstellar medium. First, we study the abundance of hydrogenation and charge states  of PAHs due to photo-ionization, photo-dissociation in the interstellar UV field, electron recombination and chemical reactions between PAH cations and H or H$_2$. For PAH cations, we find that the dehydrogenation effects are dominant. The hydrogenation state of PAHs  depends strongly on the H density, the size of the molecule and UV field. In  diffuse clouds with  low H density and normal UV radiation, PAHs containing less than 40 C are completely or strongly dehydrogenated whereas at high H density, they are normally hydrogenated. The partially dehydrogenated species dominate in intermediate density clouds. PAHs above 40 C are quite stable and are fully hydrogenated, which would favor their spectroscopic search in near IR surveys of Diffuse Interstellar Bands (DIBs).


\end{abstract}

%

\section{Introduction}

In recent years, it has been suggested that a population of large carbonaceous molecules, such as Polycyclic Aromatic Hydrocarbons (PAHs) and fullerene compounds could play a significant role in the physics and chemistry of interstellar and circumstellar environments. These molecules have been proposed as the sources of the broad infrared emission features (Duley \& Williams 1981, L\'eger \& Puget 1984, Allamandola, Tielens \& Barker 1985) and of the diffuse interstellar absorption bands (L\'eger \& d'Hendecourt 1985, van der Zwet \& Allamandola 1985, Crawford, Tielens \& Allamandola 1985).
PAHs would represent the most abundant molecules in the interstellar medium after H$_2$ and CO (L\'eger \& d'Hendecourt 1985). They could exist under various charge and hydrogenation states depending on local conditions (density, UV flux...). Environment studies of some DIBs suggest carriers with ionization potential of 10-13~eV such as PAH cations (Cami et al. 1997, Sonnentrucker et al. 1997, 1999).

The aim of this work is to propose a description of the abundance of different hydrogenation states for some small and medium PAHs $\leq$ 50C under the physical conditions present in the interstellar medium, especially in diffuse environments where UV photons strongly influence the dissociation of PAHs.

We first compare the effect of physical processes involving PAH cations and find the balance dominated by  photo-dissociation and hydrogenation processes. Then we calculate the distribution of hydrogenation states for re\-presentative PAH cations. 

\section{Physical processes of PAHs in diffuse clouds}
\vspace{-0.1cm}

We discuss here the processes involving PAHs in diffuse interstellar clouds. There are four main processes in competition: photo-dissociation following the absorption of UV photons, photo-ionization, electronic recombination and reactions between PAH cations and H or H$_2$.

\vspace{-0.1cm}
\subsection{Photo-dissociation}

Photo-dissociation takes place when the photon absorbed by the molecule has a sufficient energy to dissociate the molecule. We only take into account the dissociation of a H or H$_2$ of PAH cations.

After an UV photon is absorbed, the PAH cation is either dissociated, or stabilized by the emission of IR photons. The UV photon energy is distributed on all vibrational degrees of freedom of the molecule. When  the ener\-gy accumulated in a vibration mode becomes comparable to the C$-$H dissociation energy ($\simeq$ 4 eV for PAH cations), the molecule dissociates.

Photo-dissociation studies are often analysed using statistical models like the RRKM (Rice-Ramsperger-Kassel-Marcus) theory, where one supposes the excited molecule is quickly relaxed into the electronic fondamental state by internal conversion. Here, we use the photo-dissociation rates calculated by Allain et al. (1996) (cf. Table 1)

\subsection{Reactions between PAH cations and H or H$_2$}

According to Le Page (private communication), for small PAH cations, the Hydrogen addition rate might be proportional to that of benzene multiplied by the ratio between the number of reactive C and the total number of C. The reactive C atoms have one bond with a H. 
The addition rates for some small PAH cations are given in Table 1. The reaction rate between an ion and a neutral can be estimated by the Langevin rate ($\simeq$ 10$^{-9}$ cm$^3$ s$^{-1}$) which is larger than the kinetic reaction rate between two neutrals ($\simeq$ 10$^{-11}$ cm$^3$ s$^{-1}$).

\begin{table}
\caption[]{Table of photo-dissociation rates of PAH cations and neutral PAHs derived from Allain et al. 1996 and addition rates of PAH cations from Scott et al. and Le Page et al.}
\scriptsize
\begin{tabular}{ccccccc}
\multicolumn{1}{c}{PAH} & \multicolumn{2}{c}{cation} &
\multicolumn{2}{c}{neutral} \\
\hline
           & H loss   & H$_2$ loss & H loss   & H$_2$ loss & k$_{add,H}$ \\
           & s$^{-1}$ & s$^{-1}$   & s$^{-1}$ & s$^{-1}$ & cm$^3$s$^{-1}$ \\
\hline
\hspace{-0.3cm}Benzene    & 3.67(-9) & 2.07(-9)   & 2.89(-9) & 1.50(-9) & 2.2(-10) \\
\hspace{-0.3cm}Anthracene & 5.67(-9) & 2.13(-9)   & 3.40(-9) & 1.19(-9) & 1.6(-10) \\
\hspace{-0.3cm}Pyrene     & 3.67(-9) & 1.47(-9)   & 2.05(-9) & 7.13(-10) & 1.4(-10) \\
\hspace{-0.3cm}Coronene   & 4.33(-9) & 1.37(-9)   & 1.49(-9) & 4.55(-10) & 1.1(-10) \\
\hspace{-0.3cm}Ovalene    & 7.33(-10)& 2.0(-10)   & 1.78(-10)& 4.85(-11) & 9.6(-11) \\
\hspace{-0.3cm}PAH 50C    & 9.0(-16) & 1.4(-16)   & 2.31(-16)& 3.61(-17) & 8.8(-11) \\
\hline
\end{tabular}
\end{table}

\normalsize
\subsection{Photo-ionization} 

Photo-ionization depends on the cross section of UV photon absorption and on the local UV intensity. The photo-ionization rate is given by:

\vspace{0.2cm}
\centerline{$ {\it k}_{\rm ion(\alpha)}  = N_{\rm C} \int^{13.6 {\rm eV}}_{IP(Z,N_{\rm C})} 4 {\rm \pi}  Y_{\rm ion(\alpha)}(E)\sigma_{\rm UV abs(\alpha)}(E) F(E) dE.$}
\vspace{0.2cm}

The ionisation yield $Y_{\rm ion}$=$\sigma_{\rm ion}$/$\sigma_{\rm UV}$ is given by Verstraete et al. (1990). We take the ionisation potential {\it IP(Z,N$_{\rm C}$)} from Dartois and d'Hendecourt (1997). The UV intensity {\it F(E)} is given by Gondhalekar et al. (1980).

\subsection{Electronic recombination}

We calculate the recombination rate using the cross section of recombination for plane molecules from Spitzer (1978) and Verstraete et al. (1990). We find:

\vspace{0.2cm}
$ {{\it k}_{\rm rec} (\rm cm^3 s^{-1}) = 1.66 \times 10^{-5} \times \sqrt{\frac{100}{\it T({\rm K})}}\times \sqrt{\frac{\it N_{\rm C}}{50}}}$
\centerline{$ {+  6.3 \times 10^{-8} \times \sqrt{\frac{T({\rm K})}{100}} \times \frac{N_{\rm C}}{50}}.$}
\vspace{0.2cm}

Even though the Spitzer's approximation may be one order of magnitude lower than the recombination rate of the naphtalene cation (mesured to be 3 $\times$ 10$^{-7}$ cm$^3$ s$^{-1}$ by Abouelaziz et al. 1993), this theory may be justified for relatively large PAHs (above 30-50 C), in absence of experimental studies.
 
\subsection{Respective roles of PAH processes}

Interstellar diffuse cloud environments are characterized by low H densities (0.1$-$100 cm$^{-3}$) and an important UV flux (Gondhalekar, 1983). We suppose the temperature constant and $\simeq$ 100K. The electronic density is supposed to be equal to the C density and all C atoms are supposed to be ionized in diffuse clouds with $n_{\rm e}/n_{\rm H} = n_{\rm C}/n_{\rm H}$ $\simeq$ 1.4 10$^{-4}$. 

We now use these main processes rates to estimate the abundance of different species of PAHs under dif\-ferent charge and hydrogenation states. We suppose a stationary state. The abundances are normalized to the total abundance:

\vspace{0.2cm}
\centerline{$\sum_i [{\rm C}_m{\rm H}_i] = 1.$}
\vspace{0.2cm}

The density of each species is given by the equilibrium equation:
$$\small  \rm {\frac{d[C_{\it m}H_{\it i}^+]}{d{\it t}} = - \sum loss + \sum gains.}$$
\footnotesize
$$ \rm{\frac{d[C_{\it m}H_{\it i}^+]}{d{\it t}} = -({\it k}_{rec} {\it n}_e + {\it k}_{diss,H} +  {\it k}_{diss,H_2} + {\it k}_{add,H} {\it n}_H}$$ 
$$ \rm {+ {\it k}_{add,H_2} {\it n}_{H_2}) [C_{\it m}H_{\it i}^+] + {\it k}_{ion} [C_{\it m}H_{\it i}] + {\it k}_{diss,H} [C_{\it m}H_{\it i+1}^+] +}$$
$$ \rm{ {\it k}_{diss,H_2} [C_{\it m}H_{\it i+2}^+] + {\it k}_{add,H} {\it n}_H [C_{\it m}H_{\it i-1}^+] + {\it k}_{add,H_2} {\it n}_{H_2} [C_{\it m}H_{\it i-2}^+],}$$
\normalsize

where $i$ represents the hydrogenation state of PAH, $n_{\rm e}$, $n_{\rm H}$ and $n_{\rm H_2}$ respectively the electron, H and H$_{\rm 2}$ density, $k_{\rm rec}$ electronic recombination rate between PAH$^+$ and electron, $k_{\rm diss,H}$ and $k_{\rm diss,H_2}$ photo-dissociation rates after UV photon absorption with ejecting a H or H$_{\rm 2}$, $k_{\rm add,H}$ and $k_{\rm add,H_2}$ H and H$_{\rm 2}$ addition rates of PAH cations, $k_{\rm ion}$ ionisation rate of neutral PAH.

We resolve these linear equations using a IDL (Interactive Data Language) programme.  
We find that the PAHs tend to be dehydrogenated in physical conditions such as in diffuse cloud environments where the Hydrogen density is about 0.1$-$100 cm$^{-3}$, the UV intensity is assumed to be constant (Gondhalekar intensity) and the temperature is about 100K.
In what follows, we will then consider the two dominant processes concerning hydrogenation states of PAH cations: photo-dissociation and hydrogenation. 

\section{Hydrogenation states of PAH cations}

The hydrogenation state of PAHs depends on two factors: first, the hydrogenation rate, which is function of Hydrogen and PAH cation density and second, the photo-dissociation rate which strongly depends on the size of the molecules and the UV intensity. The equilibrium state is a balance between the photo-dissociation rate due to UV photon absorption and the hydrogenation rate. We will present a simple model which takes only these two processes into account.

\vspace{0.2cm}
$\rm { [C_{\it m}H_{\it n}^+] \leftrightarrow [C_{\it m}H_{\it n-1}^+] \leftrightarrow  [C_{\it m}H_{\it n-2}^+] \leftrightarrow ... }$ 
\centerline{$\rm {\leftrightarrow [C_{\it m}H_2^+] \leftrightarrow [C_{\it m}H_1^+] \leftrightarrow [C_{\it m}^+]}$}

\vspace{0.2cm}
\centerline{$ \rm {C_{\it m}H_{\it n} + UV \rightleftharpoons^{diss}_{add} C_{\it m}H_{\it n-1} + H}$}

\vspace{0.2cm}
The goal of the model is to estimate the distribution of hydrogenation states for a given PAH cation as a function of environment parameters in the interstellar medium clouds.

\vspace{-0.2cm}
\subsection{Case 1: with constant weighting of $k_{\rm add}$ vs H cover}

We assume for $1 < i < n$ the balance:

\vspace{0.2cm}
\centerline{$ \rm {\frac{[C_{\it m}H_{\it i}] [\Phi_{\nu}]}{[C_{\it m}H_{\it i-1}] [H]} = \frac{{\it k}_{add}}{{\it k}_{diss}},}$}
\vspace{0.2cm}

where $k_{\rm add}$ and $k_{\rm diss}$ represent the hydrogenation rate and the photo-dissociation rate, $\Phi_{\rm \nu}$ and [H] are the local UV flux and the H density.

In this case, we assume that the probability of capturing or loosing a H is the same for various hydrogenation states of PAHs. Then in the second case, we will assign  a weight to each hydrogenation state.

Let us put  $\rm {{\it x}_H  = \frac{{\it k}_{add}}{{\it k}_{diss}} \times \frac{[H]}{\Phi_{\nu}}.}$ One has: 

\vspace{0.2cm}
\centerline{$\rm {\frac{[C_{\it m}H_{\it i}]}{[C_{\it m}H_0]} = ({\it x}_H)^{\it i}.}$}
\vspace{0.2cm}

The number of molecules normalized by the total abundance is:

\vspace{0.2cm}
\centerline{$\rm {\sum_{\it i=0}^{\it n}[C_{\it m}H_{\it i}] = \sum_{\it i=0}^{\it n} ({\it x}_H)^{\it i}[C_{\it m}H_0] = 1.}$}
\vspace{0.2cm}

The abundance for each hydrogenation state is:

\vspace{0.2cm}
\centerline{$\rm {[C_{\it m}H_{\it i}] = ({\it x}_H)^{\it i} [C_{\it m}H_0] = ({\it x}_H)^{\it i} \frac{(1-{\it x}_H)}{(1-({\it x}_H)^{\it i+1})},}$}

\begin{figure}
\centerline{\psfig{file=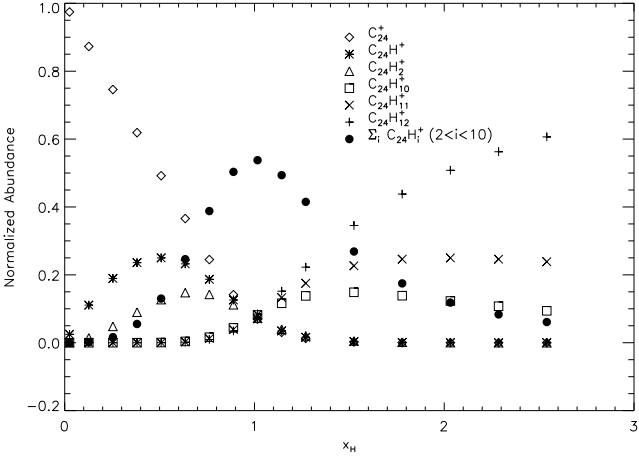,height=6.5cm}}
\caption{Abundance of hydrogenation states of the coronene cation as a function of $x_{\rm H}$. Three  phases are present with increasing the H density. At $x_{\rm H}$ $<$ 0.8, strongly dehydrogenated species dominate. At $x_{\rm H}$ around 1 partially dehydrogenated species dominate. At $x_{\rm H}$ $>$ 1.5, the coronene cation is completely hydrogenated.}\label{corox_H}
\end{figure}

$x_{\rm H}$ = 1 corresponds to the maximum of intermediate species for $n_{\rm H}$ = 67, 50, 74, 14, 1.7 $\times$ 10$^{-2}$, 2 $\times$ 10$^{-5}$ cm$^{-3}$ respectively for the anthracene cation C$_{14}$H$_{10}^+$, the pyrene cation C$_{16}$H$_{10}^+$, the coronene cation C$_{24}$H$_{12}^+$, the ovalene cation C$_{32}$H$_{14}^+$, the PAH 40C cation and the PAH 50C cation. We note the decrease of n$_{\rm H}$ with PAH sizes, except for coronene which has a high dissociation rate because of its particularly symmetric configuration, which limits radiative relaxation emission of infra-red vibrational modes. 

\vspace{-0.2cm}
\subsection{Case 2: variable weighting of $k_{\rm add}$ vs H cover}

We assume: \hspace{0.5cm} $ \rm \frac{[C_{\it m}H_{\it i}] [\Phi_{\nu}]}{[C_{\it m}H_{\it i-1}] [H]} = \frac{{\it k}_{add}}{{\it k}_{diss}} \times \frac{\it n-(i-1)}{\it n}.$

\vspace{0.2cm}
In this case, each hydrogenation state of PAH cations is weighted by a factor $\frac{n-i}{n}$ in order to take into account the number of sites available for adding a H atom.

The abundance for each hydrogenation state is given by:

\vspace{0.2cm}
$\frac{[{\rm C}_m{\rm H}_i]}{[{\rm C}_m{\rm H}_0]} = (x_{\rm H})^i \times \frac{n-(i-1)}{n} \times ....\times \frac{n-1}{n} \times \frac{n}{n}.$
\vspace{0.2cm}

The number of molecules normalized by the total abundance is:

\vspace{0.2cm}

\centerline{\small $\sum_{i=0}^{n}[{\rm C}_m{\rm H}_i] = [{\rm C}_m{\rm H}_0] \left( 1 + \sum_{i=1}^{n}  (x_{\rm H})^i \Pi_{j=1}^{j=i} \frac{n-(j-1)}{n} \right) = 1.$}
\vspace{0.2cm}

We have calculated the balance between hydrogenation and photo-dissociation. The results are given in Fig.2.

\section{Discussion}

\subsection{Dependence on H density and on molecule size}

The hydrogenation state strongly depends on the H density. Two processes are competing: that of the UV radiation which leads to the dehydrogenation of  PAH cations and the reaction between PAH cations and H which favors hydrogenated states.

\begin{figure*}
\hspace{-0.5cm}
\begin{tabular}{cc}
\psfig{file=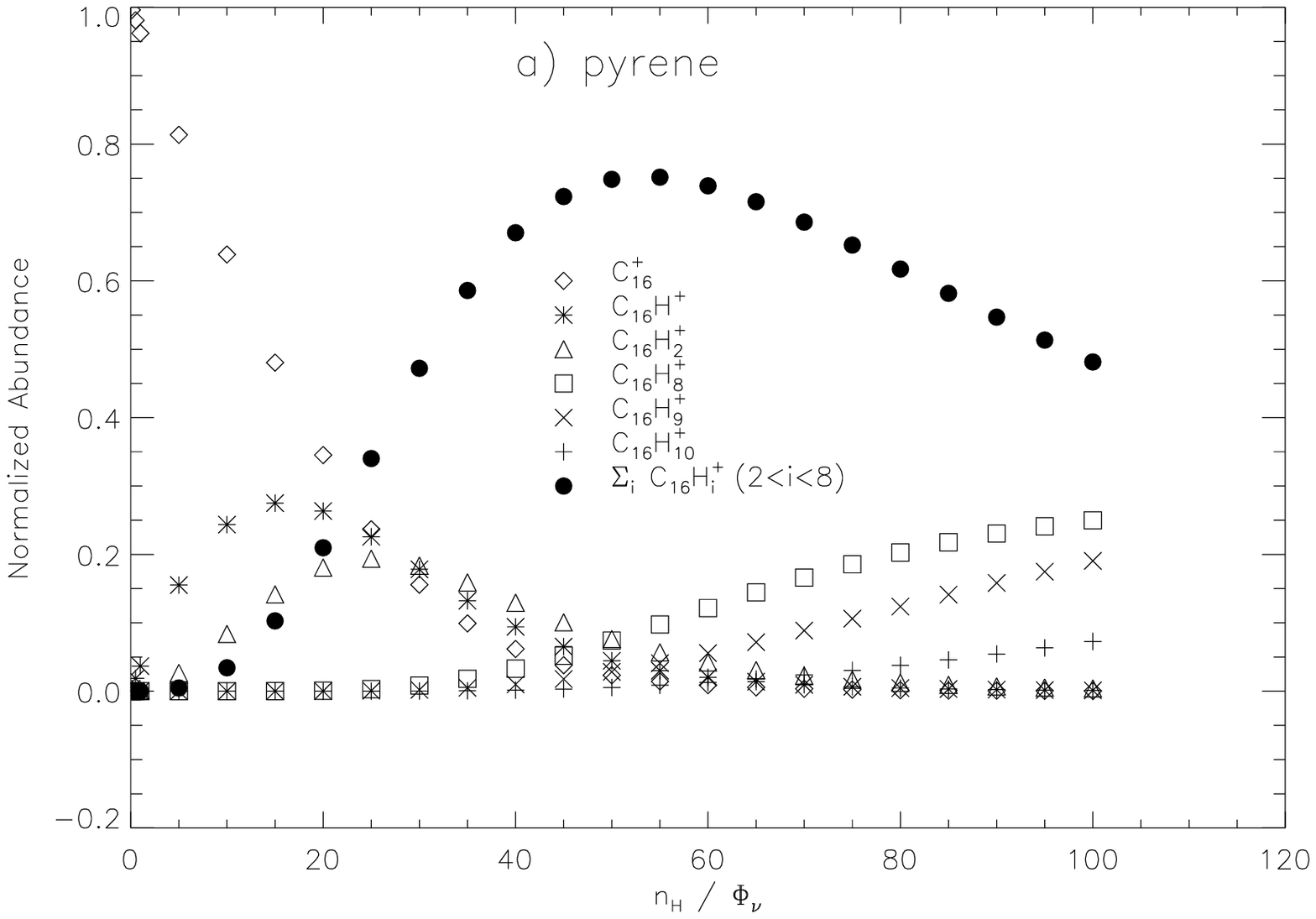,height=6.4cm} &
\psfig{file=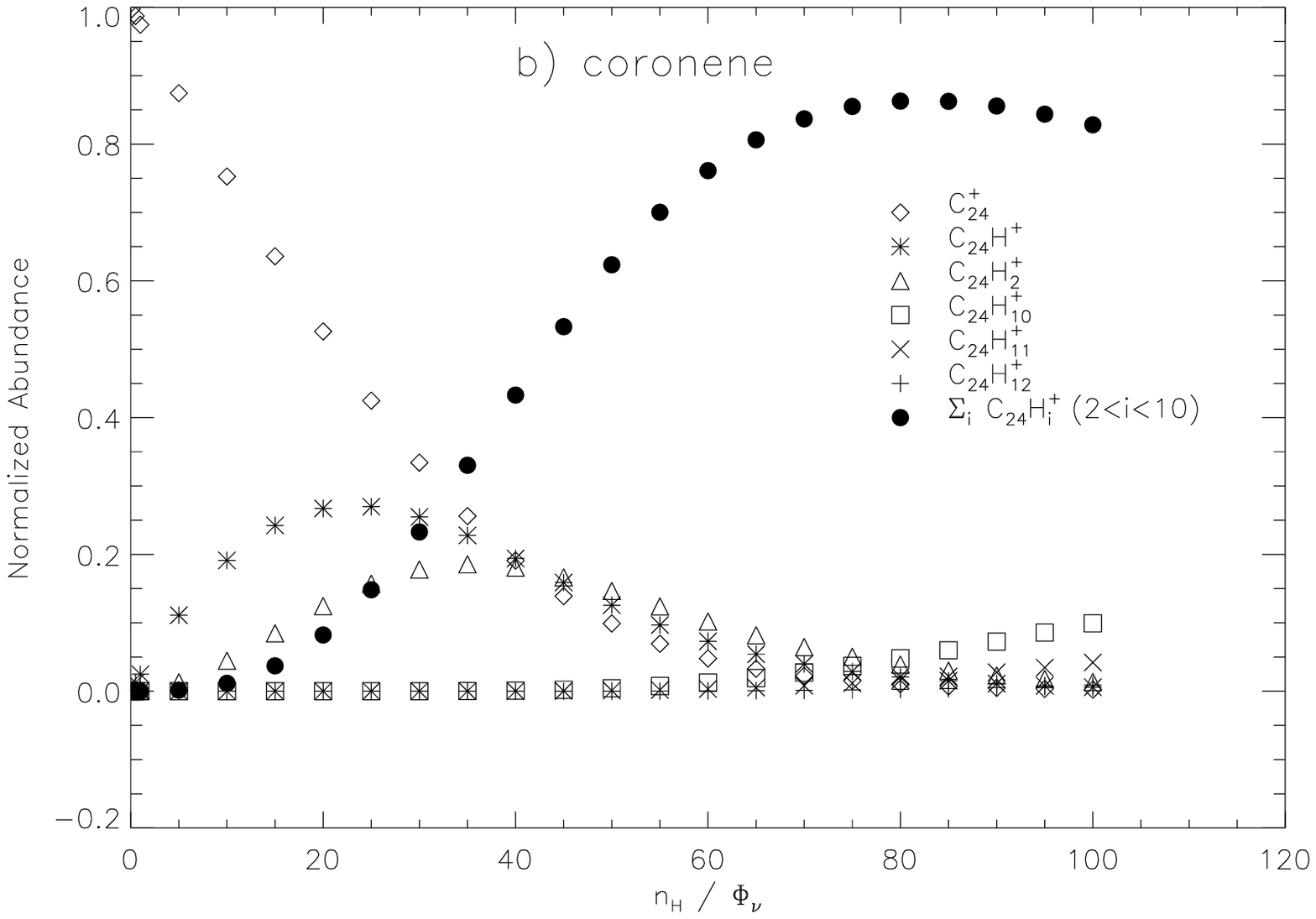,height=6.4cm} \\
\psfig{file=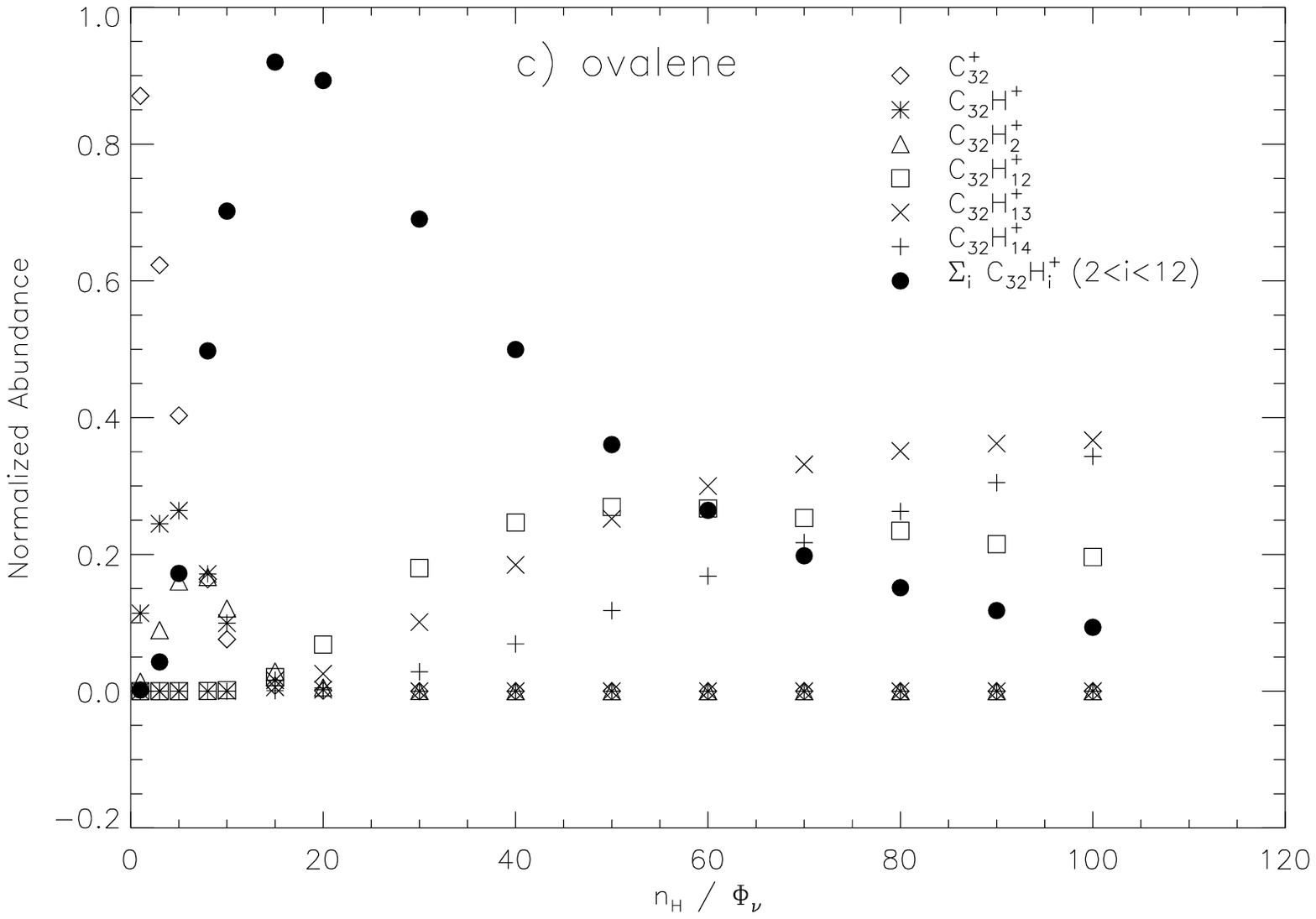,height=6.4cm} &
\psfig{file=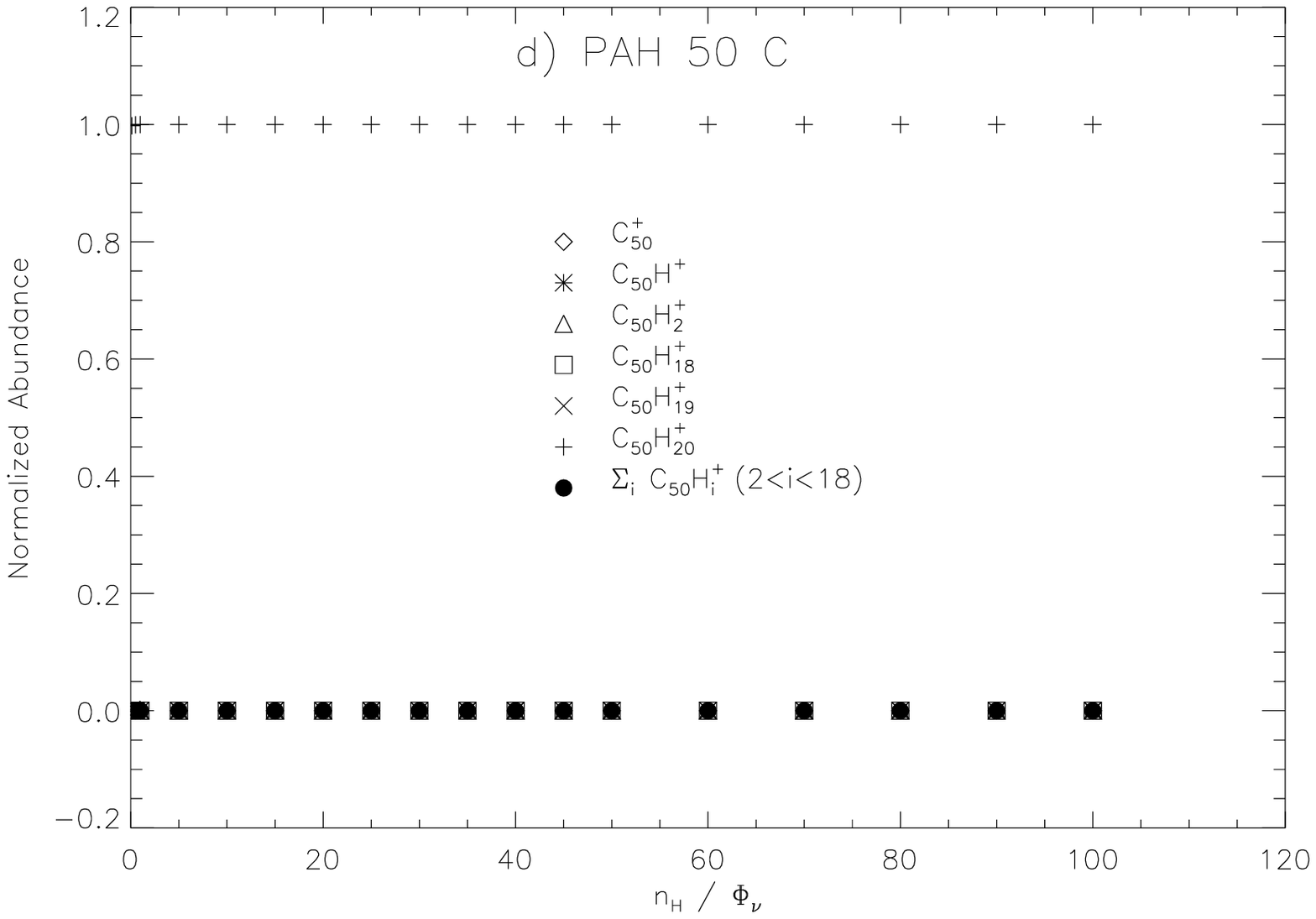,height=6.4cm} \\
\end{tabular}
\label{coro}
\label{ova}

\caption{Different hydrogenation states of a) the pyrene cation C$_{16}$H$_{10}^+$, b) the coronene cation C$_{24}$H$_{12}^+$, c) the ovalene cation C$_{32}$H$_{14}^+$ and d) the PAH 50C cation as a function of H density $n_{\rm H}$ in the presence of UV (Gondhalekar spectrum multiplied by a constant factor $\it \Phi_\nu$). At low densities, strongly dehydrogenated species dominate whereas at $n_{\rm H}$ $>$ 20, 30, 5 cm$^{-3}$ respectively for the pyrene cation, the coronene cation, the ovalene cation, partially dehydrogenated species are important. The normal hydrogenated PAH 50C cation is stable over a large range of densities.}

\label{pyr}
\label{PAH50}

\end{figure*}
As we can see in the figures 2, three phases exist for each hydrogenation state of PAHs except for PAHs $>$ 40C. The UV intensity is assumed to be constant while the H density varies. At low density, PAH cations are completely or strongly dehydrogenated but at high density, they are normally hydrogenated. At intermediate density, PAH cations are partially dehydrogenated. 

For PAH cations $\geq$ 50C, the completely hydrogenated state is favored (cf. Fig.\ref{PAH50}d). This can be explained by the fact that large molecules have more degrees of freedom available to store the excess energy brought by the UV photon, which decreases the photodissociation rate.

\subsection{Search for PAHs and link with DIBs}

Small PAHs $<$ 30C will be difficult to search because their abundance is distributed over a number of hydrogenation states. Their superimposed absorptions can contribute to the envelope of quasi continuum extinction in addition to dust due to a large number of states and isomers. PAHs $>$ 40C are stable and fully hydrogenated. Their signature can be searched in the fundamental band transition expected in the far red  and near infra-red around 1 $\mu$m (Foing\- \& Ehrenfreund 1994). Rotational contours for such PAH $>$ 40C would lead to FWHM broadening of 2-4 cm$^{-1}$ depending on the exact size and geometry of the molecule (Ehrenfreund \& Foing 1996). These transitions can be searched in near IR surveys of DIBs.
Laboratory spectroscopy and dissociation and hydrogenation rates are also needed for guiding the search for PAHs in space.

\begin{acknowledgements}

We thank the referee V. Le Page for helpful comments which have substantially improved the paper.

\end{acknowledgements}

\end{document}